\documentclass[]{spie}  %>>> use for US letter paper
\usepackage{amsmath,amsfonts,amssymb}
\usepackage{graphicx}
\usepackage[colorlinks=true, allcolors=blue]{hyperref}

\title{High contrast imaging with ELT/METIS: The wind driven halo, from SPHERE to METIS}

\author[a,*]{Cantalloube~F.}
\author[b]{Absil~O.}
\author[a]{Bertram~T.}
\author[a]{Brandner~W.}
\author[b]{Delacroix~C.}
\author[a]{Feldt~M.}
\author[c]{Kenworthy~M.~A.}
\author[a]{Kulas~M.}
\author[f]{Milli~J.}
\author[d]{Neureuther~P.}
\author[b]{Orban de Xivry~G.}
\author[e]{Pathak~P.}
\author[c]{Por~E.}
\author[a]{Scheithauer~S.}
\author[a]{Steuer~H.}
\author[a]{van Boekel~R.}
\affil[a]{Max Planck Institut f\"ur Astronomie, K\"onigstuhl 17, D-69117, Heidelberg, Germany}
\affil[b]{Space sciences, Technologies and Astrophysics Research (STAR) Institute, Universit\'e de Li\`ege}
\affil[c]{Leiden Observatory, Leiden University, P.O. Box 9513, 2300 RA Leiden, The Netherlands}
\affil[d]{Institute for System Dynamics, University of Stuttgart, Germany}
\affil[e]{ESO, Karl-Schwarzschild-Strasse 2, 85748 Garching, Germany}
\affil[f]{ESO, Alonso de C\'ordova 3107, Vitacura, Casilla 19001, Santiago, Chile}

\authorinfo{Further author information: (Send correspondence to F.C.)\\F.C.: E-mail: cantalloube@mpia.de, Telephone: 49 622 123 1234}

\begin{document} 
\maketitle

\begin{abstract}
METIS is one of the three first-light instruments planned for the ELT, mainly dedicated to high contrast imaging in the mid-infrared. On the SPHERE high-contrast instrument currently installed at the VLT, we observe that one of the main contrast limitations is the wind driven halo, due to the limited AO running speed with respect to the atmospheric turbulence temporal evolution. From this observation, we extrapolate this signature to the ELT/METIS instrument, which is equipped with a single conjugated adaptive optics system and with several coronagraphic devices. By making use of an analytic AO simulator, we compare the amount of wind driven halo observed with SPHERE and with METIS, under the same turbulence conditions.
\end{abstract}


\section{INTRODUCTION}
\label{sec:intro}  
High-contrast imaging (HCI) is one of the main challenges for future astronomical observation, especially to study exoplanet formation and evolution within their environment. %to detect and characterize forming planets within disks, study the atmospheric physics and dynamics of evolved planets
In the context of the European Extremely Large Telescope (ELT) that is currently under construction at Cerro Armazones (Chile), all of the three first light instruments hold a high-contrast imaging mode. To achieve the contrast required for exoplanet detection, they are all equipped with specific coronagraphic technologies, working under adaptive optics (AO) correction, delivering images on which advanced post-processing techniques will be applied to further carve out the contrast.

\begin{figure}[ht]
   \begin{center}
   \begin{tabular}{c} %% tabular useful for creating an array of images 
   \includegraphics[scale=0.25]{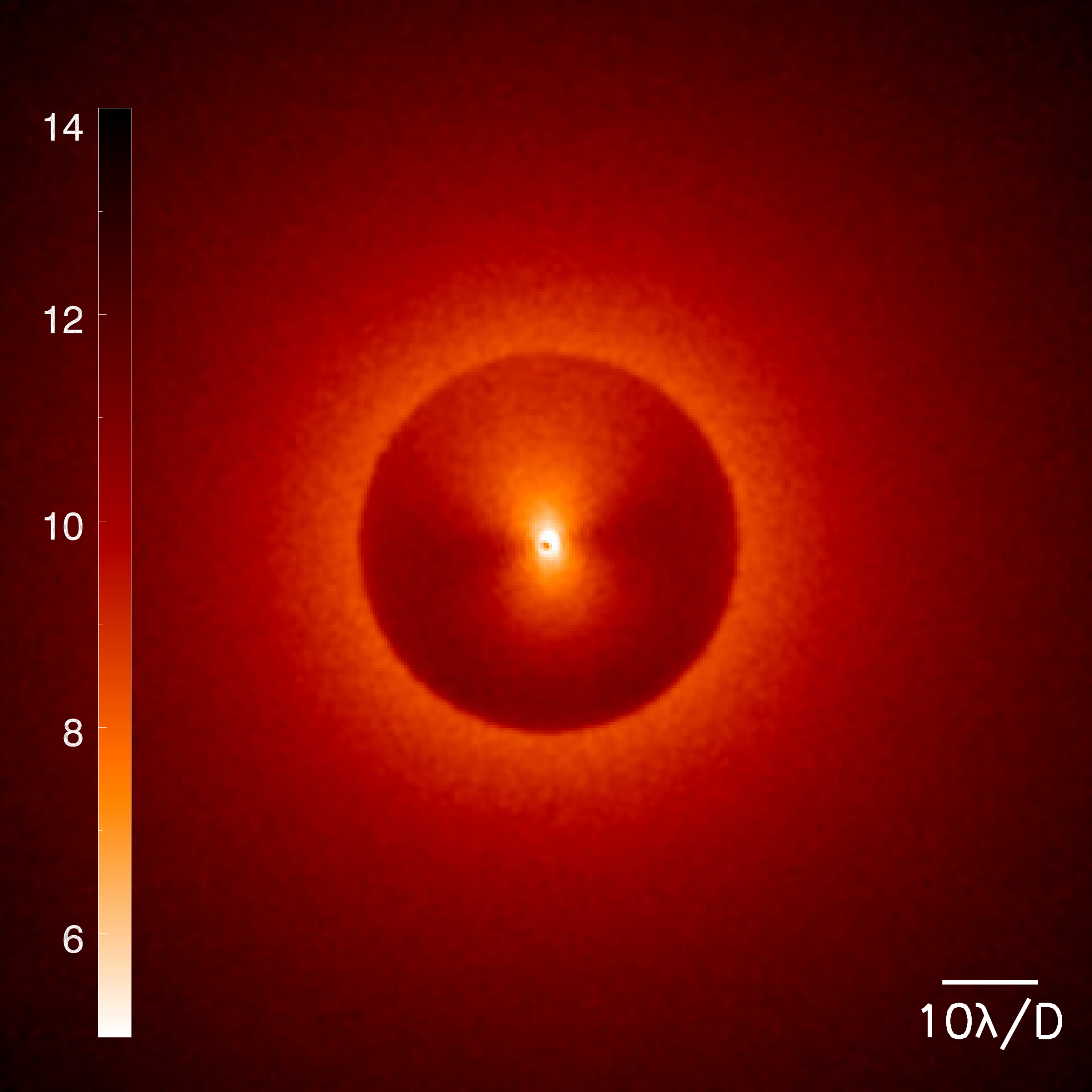}
   \includegraphics[scale=0.25]{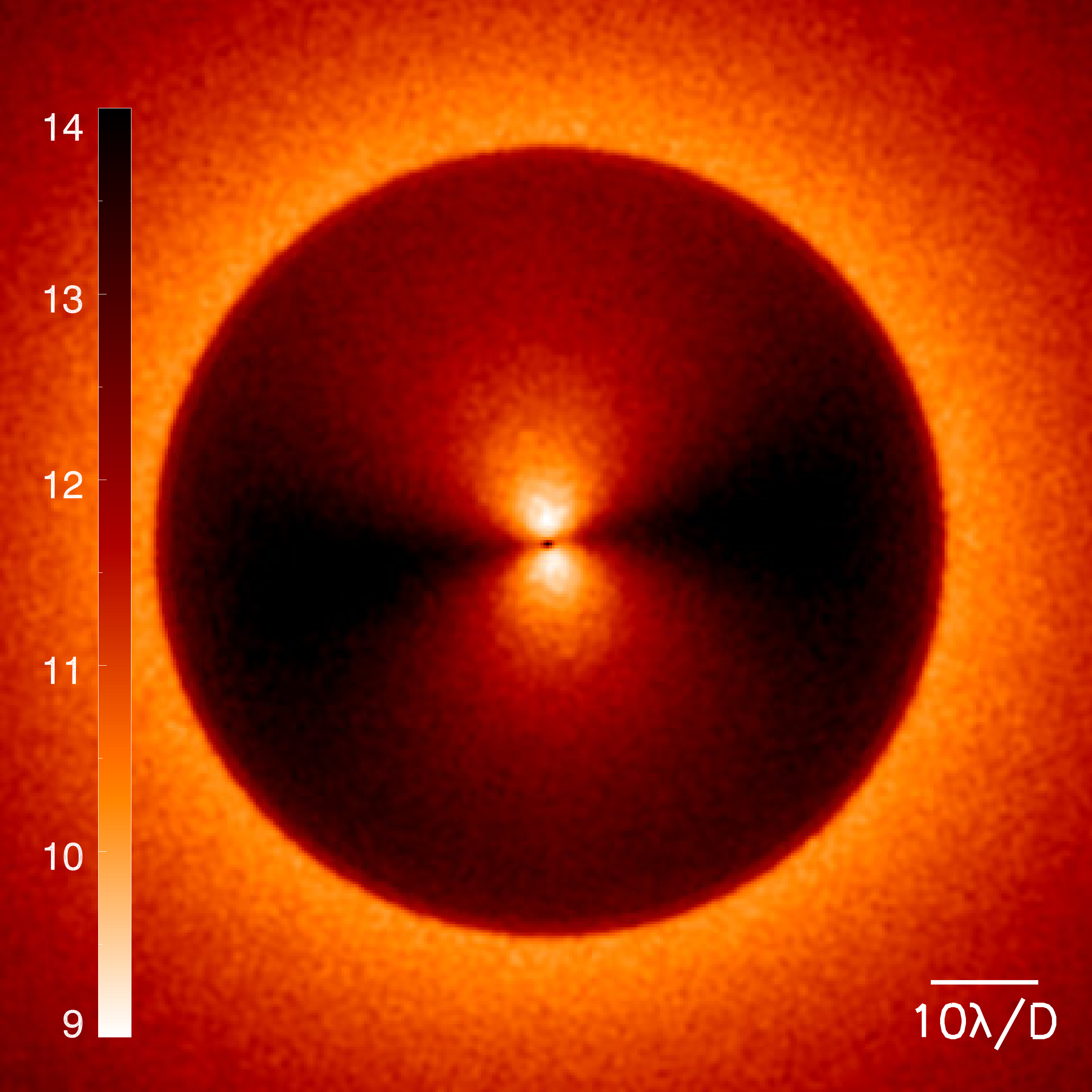}
   \end{tabular}
   \end{center}
   \caption[example] 
   {\label{fig:intro} 
Illustration of the wind driven halo in high-contrast images. Simulations of SPHERE ideal coronagraphic images in H-band (left) and of METIS ideal coronagraphic images in L-band (right), both accounting only for the AO residuals under the same turbulence conditions. The images are cropped to $384\times 384~\mathrm{pixels}$ and the color scale shows the contrast in magnitude (logarithmic scale).}
\end{figure}

The Mid-Infrared ELT Imager and Spectrograph, METIS\cite{brandl2018status, brandl2008metis}\footnote{\url{http://metis.strw.leidenuniv.nl/}}, is one of the three first light instruments of the ELT whose main science goal is to observe exoplanets and circumstellar disks\cite{quanz2015sciencecases}. It is working in the thermal infrared (from L-band, at $3.7~\mathrm{\mu m}$ to Q-band, at $19~\mathrm{\mu m}$) and equipped with a single conjugated adaptive optics (SCAO) system\cite{bertram2018scao,hippler2019scao} analyzing the atmospheric turbulence with a pyramid wavefront sensor working in the K-band ($2.2~\mathrm{\mu m}$), and with five types of coronagraphs\cite{kenworthy2018reviewcoro} designed to reach a L-band contrast of $3\times10^{-5}$ at $5~\mathrm{\lambda/D}$ (requirement) to $1\times10^{-6}$ at $2~\mathrm{\lambda/D}$ (goal). METIS has recently gone through its preliminary design review (PDR) in May 2019, and its first light at the ELT is due in 2026.  
Under good observing conditions, the high-contrast capabilities of the latest generation of HCI instruments are limited by non-common path aberrations (NCPA) resulting in the presence of bright speckles in the images hindering the presence of planetary companions. However, other limitations exist due to the structure of the telescope or due to turbulence conditions interacting with the whole instrument\cite{CantalloubeMsgr}. 
Among those limitations, the wind driven halo (WDH) is an elongation of the point spread function (PSF) along the direction of the equivalent wind at the telescope pupil, that is revealed as a butterfly-shaped pattern in the high-contrast images. This elongation occurs when the atmospheric turbulence is varying too fast to be fully corrected by the AO system. This effect has been clearly seen on the latest generation of HCI instruments such as VLT/SPHERE\cite{beuzit2019sphere,Cantalloube2018} or Gemini-S/GPI\cite{Macintosh2008,Madurowicz2019}. According to turbulence profiling data from Stereo-SCIDAR measurements\cite{osborn2018scidar}, this situation happens up to 40\% of the observing time at the Paranal observatory. The WDH is illustrated in Fig.~\ref{fig:intro} showing simulations with an ideal coronagraph\cite{cavarroc2006idealcoro,sauvage2010analyticalcoro} for a SPHERE-like system (left) and METIS-like system (right), in which only the contribution from AO residuals is taken into account.

After introducing the origin of the WDH in the following section, we present simulated images for a SPHERE-like instrument and for a METIS-like instrument to evaluate the impact of the WDH in the raw contrast reached in the delivered images when going from one to the other system. % (Sect.~\ref{sec:main)}).

%%%%%%%%%%%%%%%%%%%%%%%%%%%%%%%%%%%%%%%%%%%%%%%%%%%%%%%%
\section{The wind driven halo}
\label{sec:wdh}  
The wind driven halo (WDH) is due to the interaction between the turbulence decorrelation speed, parametrized by the turbulence coherence time, and the finite speed at which the AO loop is running, inducing delays in the correction, provoking the so-called AO servolag (or temporal bandwidth) error.

\subsection{The adaptive optics delay}
For an AO system running in closed loop with a classical integrator law, the AO servolag error is due to the delay between the measurement of the turbulent phase at the wavefront sensor (WFS) detector and the effective correction by the deformable mirror (DM). It is therefore dependent on the WFS integration time, the real time computer latency and the time the DM takes to be fully in shape. The AO loop frequency is the frequency at which a new turbulent phase is integrated at the WFS detector. The shape of the DM is therefore refreshed at the same frequency. For the VLT/SPHERE instrument, the routine AO loop frequency used for bright targets is $1380~\mathrm{Hz}$ (set by the integration time needed at the WFS detector) and the AO-loop delay has been measured at $2.3$ frames. For the ELT/METIS instrument, the AO loop frequency is limited to $1000~\mathrm{Hz}$ (set by the maximum running frequency of the M5 DM) and, in the following, we assume a frame delay of $2$.% with a scalar gain. 

If the atmospheric turbulence is varying faster than the AO loop delay, the corrected phase (or AO residual phase) will show strong low spatial frequency residuals.

\subsection{Temporal variation of the atmospheric turbulence}
The turbulence coherence time characterizes the temporal evolution of the atmospheric turbulence. It is defined\cite{Roddier1981} as $\tau_0 = 0.314 \; r_0 / \mathbf{v_{eq}}$, where $r_0$ is the Fried parameters (defined at $500~\mathrm{nm}$ and at zenith) and $\mathbf{v_{eq}}$ is the equivalent wind velocity at the telescope pupil, that is to say the integrated wind speed weighted by the index structure constant ($C_n^2$)\cite{Roddier1981}. At the Paranal observatory, the median coherence time is $4.6~\mathrm{ms}$ based on MASS-DIMM historical data from April 2016 to April 2019. At the Cerro Armazones, the median coherence time is $3.5~\mathrm{ms}$\cite{vernin2008elt}.

When the turbulence coherence time is shorter than or close to the AO loop delay, the AO corrected phase shows strong low spatial frequency phase residuals along the equivalent wind direction. When propagated to the focal plane, those directional phase residuals provoke an elongation of the PSF, along the direction of the equivalent wind, at a contrast of up to $10^{-4}$. This scattered light, forming the WDH, is therefore revealed only when using a coronagraphic device that cancels most of the on-axis starlight (see Fig.~\ref{fig:coronagraph}).

When modeling the atmosphere as a finite number of turbulent layers located at different altitudes, under the Taylor hypothesis (the frozen flow hypothesis), the main temporal variation is due to the wind that translates each layer at a different speed and direction. Since high equivalent wind velocities are the main responsible for the WDH, the WDH appears most of the time in the presence of strong jet stream, a thin layer of a few kilometers width, located at about $12~\mathrm{km}$ above sea level (near the tropopause) and whose speed can go from $20$ up to $50~\mathrm{m/s}$, flowing west to east. We indeed observed on VLT/SPHERE\cite{beuzit2019sphere} high-contrast images that the WDH is almost always aligned with the direction of the subtropical jet that crosses the Paranal observatory location and is tightly correlated to the presence of strong jet stream. Using Gemini-S/GPI\cite{Macintosh2008} (located at Cerro Pachon, Chile) high-contrast images, a team correlated the WDH direction to the different wind layer direction and came to the same conclusions\cite{Madurowicz2018SPIE}. 

In addition, we observed in high-contrast images delivered by the latest generation of HCI instrument (equipped with both extreme AO, XAO, and advanced coronagraph designs) that the WDH is asymmetric, one wing of the butterfly-shaped pattern being wider and brighter than the other. We recently found out that this asymmetry is due to interference between delayed phase error and amplitude errors\cite{Cantalloube2018}. The delayed phase error is caused by the AO servolag and the amplitude errors are caused by the scintillation effect that are phase errors translating into amplitude error due to Fresnel propagation through the atmosphere. Even though the scintillation effect is negligible for 8-m telescopes working in the near infrared (contrast of more than $10^{-6}$), this is still enough to create this asymmetry that any type of coronagraph would reveal. This asymmetry increases with the scintillation, that is to say with smaller telescope diameter, larger wavelength and lower observing site (higher atmospheric layers). As this asymmetry is due to interference between the delayed phase error and the amplitude errors, it increases when the correlation between delayed phase error and amplitude errors increase, that is to say with slower equivalent wind velocity and/or smaller AO loop delay.

\subsection{Power spectral density of the AO residuals}
\begin{figure} [ht]
\begin{center}
\begin{tabular}{c} %% tabular useful for creating an array of images
\resizebox{\hsize}{!}{\includegraphics{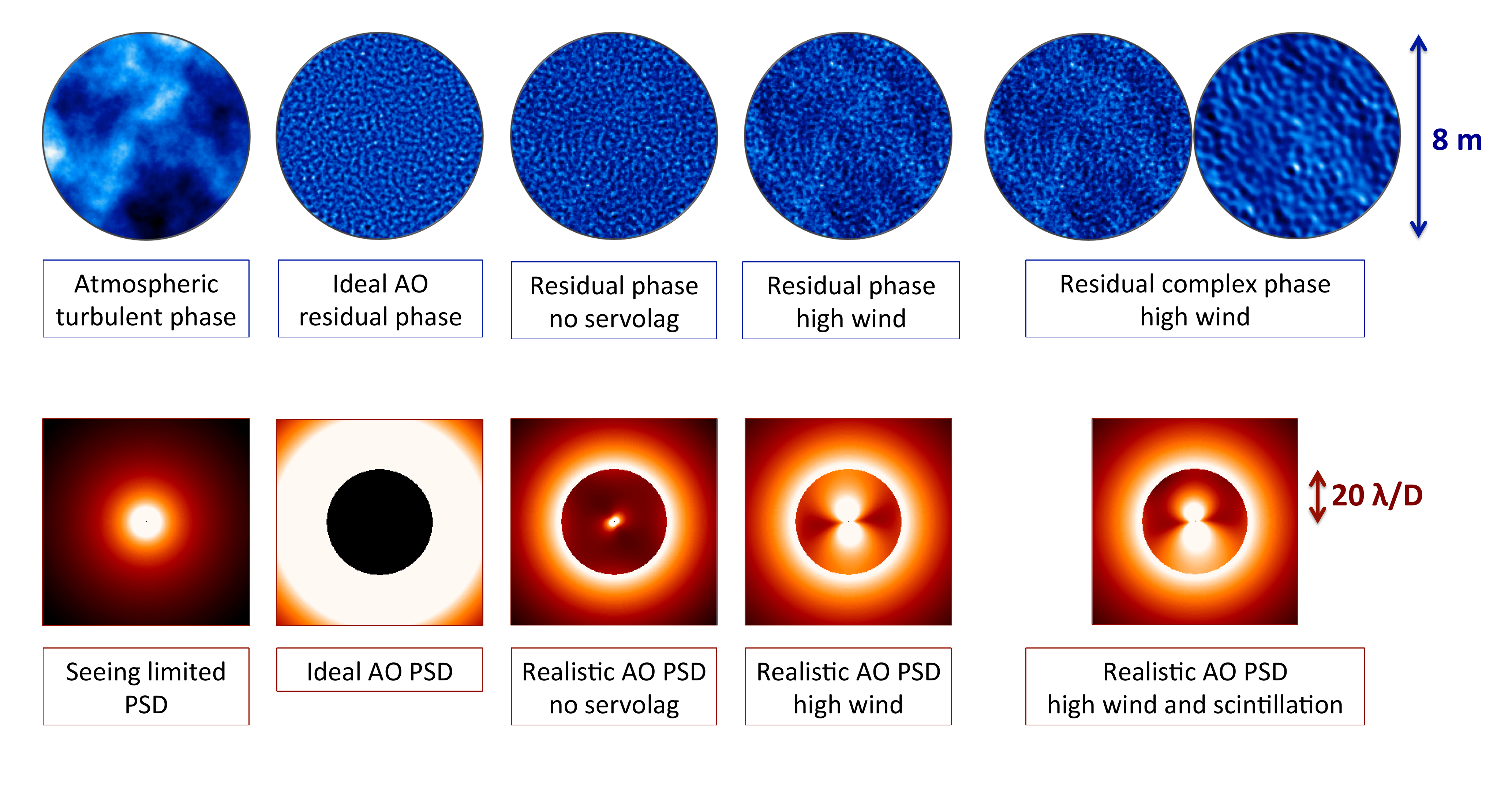}}
\end{tabular}
\end{center}
\caption{\label{fig:psd} Illustration of the AO servolag error for a SPHERE-like system, with five different scenarios (from left to right): Without AO correction (seeing limited), with a perfect AO correction only limited to the finite density of the DM (fitting error), AO residuals without the servolag error, AO residuals with servolag error and strong equivalent wind speed and AO residuals additionally taking into account the interference with amplitude errors. Top row (blue): Phases projected on the $8~\mathrm{m}$ diameter telescope. 
Bottom row (red): Corresponding power spectral densities showing the spatial frequencies distribution of the AO residual terms.}
\end{figure}

To illustrate the explanations from previous sections, Fig.~\ref{fig:psd} shows the phase (upper row) and the corresponding power spectral densities (lower row) for a VLT/SPHERE-like XAO system observing in H-band (centered at $1.6~\mathrm{\mu m}$). From left to right, Fig.~\ref{fig:psd} shows the case without AO correction (seeing limited), with an ideal AO correction (only limited by the fitting error due to the limited number of DM actuators), the case without AO servolag error (showing other AO errors such as chromaticity, anisoplanetism, aliasing and noise propagation), and the cases with servolag with or without its interference with amplitude errors. 

\subsection{Effect of the AO servolag error in high-contrast imaging}
\begin{figure}[h]
   \begin{center}
   \begin{tabular}{c} %% tabular useful for creating an array of images 
   \includegraphics[scale=0.5]{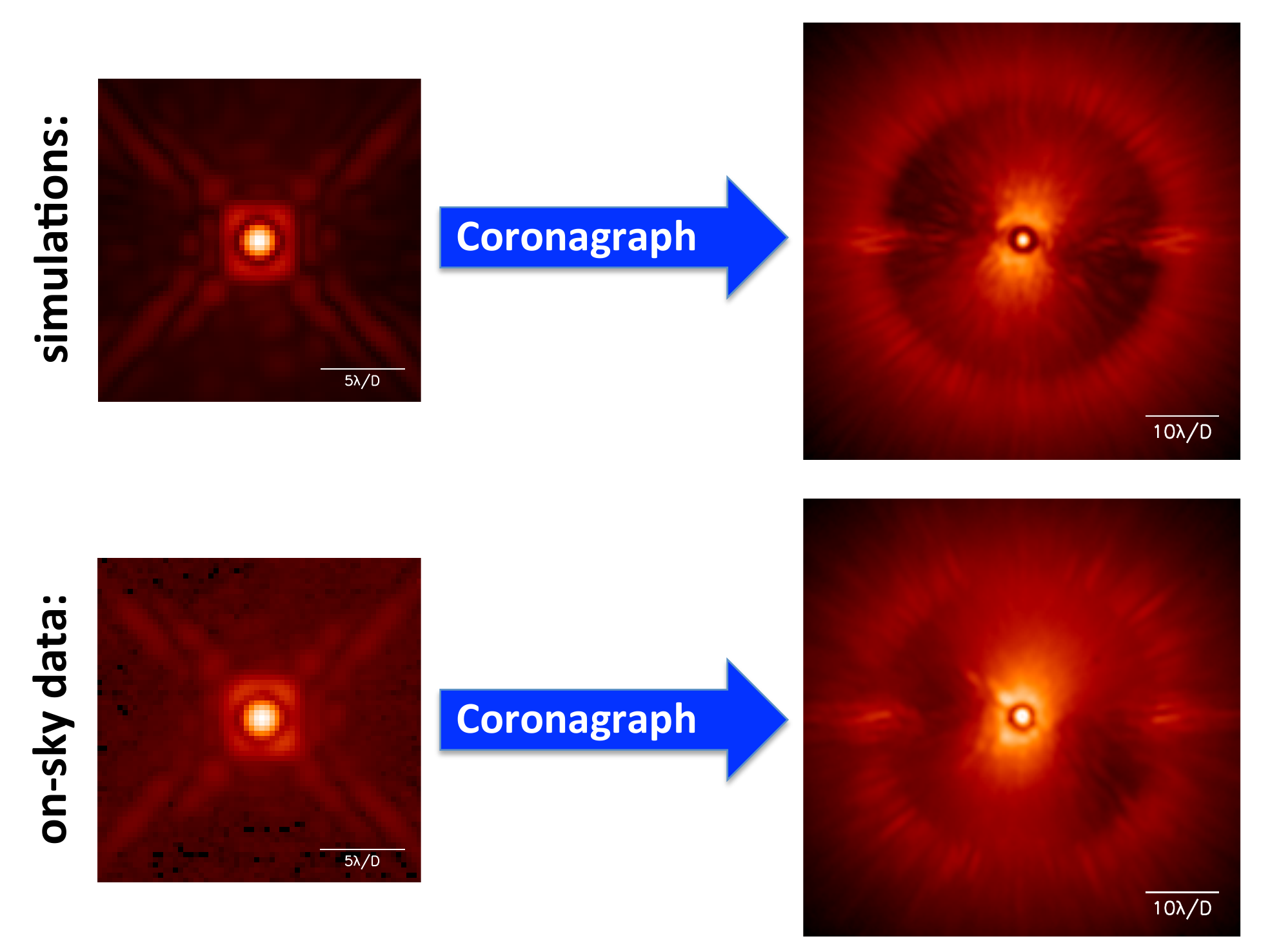}
   \end{tabular}
   \end{center}
\caption{\label{fig:coronagraph} Illustration of the effect of the coronagraph revealing the wind driven halo in the in AO-corrected images. 
Top: Simulations of SPHERE images in broadband-H (centered around $1.6~\mathrm{\mu m}$) without (left) and with (right) its APLC coronagraph.
Bottom: On-sky images from SPHERE in broadband-H without (left) and with (right) its APLC coronagraph.}
\end{figure}

The contribution of the AO servolag error in the image is expected to be $10^4$ fainter than the star. Consequently, without coronagraph, this extension of the PSF is completely hidden by the starlight. However, when using a coronagraph coupled with extreme AO, a raw contrast of $10^{-4}$ can be reached in the AO-corrected area, thus revealing the wind driven halo very clearly in the image. Figure~\ref{fig:coronagraph} shows the example of simulated (top row) and on-sky images (bottom row) of the SPHERE instrument that is equipped with an XAO and an apodized Lyot coronagraph\cite{soummer2011aplc,martinez2009aplc} (APLC): the non-coronagraphic image (left column) shows no hint of the PSF elongation whereas the coronagraphic image (right column) reveals clearly the wind driven halo. 

As shown in Fig.~\ref{fig:coronagraph}, the simulations of SPHERE images are consistent with the obtained on-sky images. In the following we therefore extrapolate from a SPHERE-like system to a METIS-like system to show how the raw contrast reached in the resulting images is affected by the WDH. 

%%%%%%%%%%%%%%%%%%%%%%%%%%%%%%%%%%%%%%%%%%%%%%%%%%%%%%%%
\section{From VLT/SPHERE to ELT/METIS}
\label{sec:main}
In this section, the effect of the wind driven halo observed in SPHERE images is predicted for a METIS-like system to verify how this error term will finally affect the raw contrast. To do so, we rely on analytical AO simulations\cite{Jolissaint2006} and, in order to focus on the effect of the wind driven halo only, we only take into account the AO residuals in the image (i.~e. we ignore NCPA or other exogenous error terms). From the residual phase screens produced, we simulate ideal coronagraphic images so that we are not biased by the type of coronagraph used as the coronagraphs designed for SPHERE and METIS are quite different and therefore show different inner working angle, response and limitations. 

\begin{figure}[ht]
   \begin{center}
   \begin{tabular}{c} %% tabular useful for creating an array of images 
   \resizebox{\hsize}{!}{\includegraphics{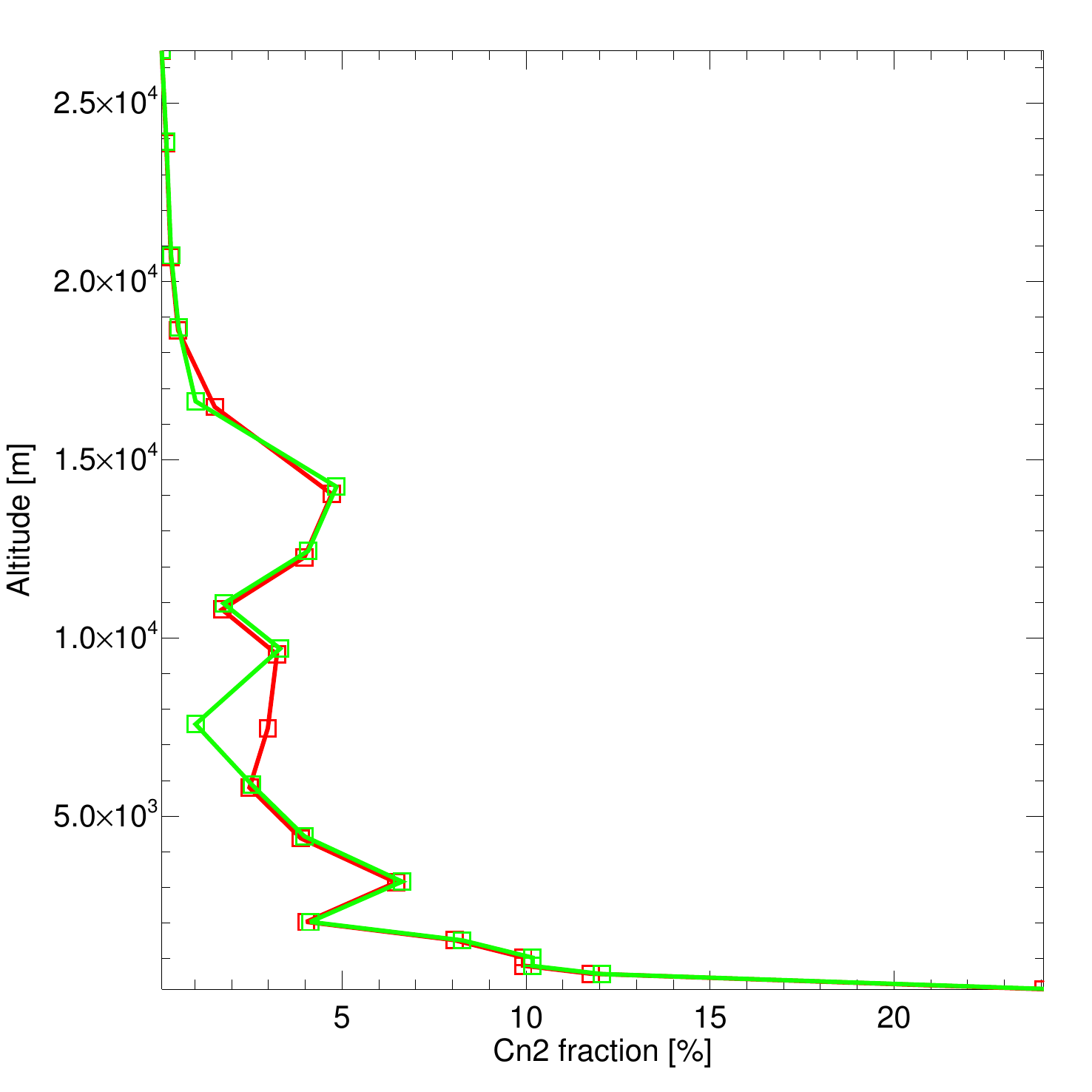}\includegraphics{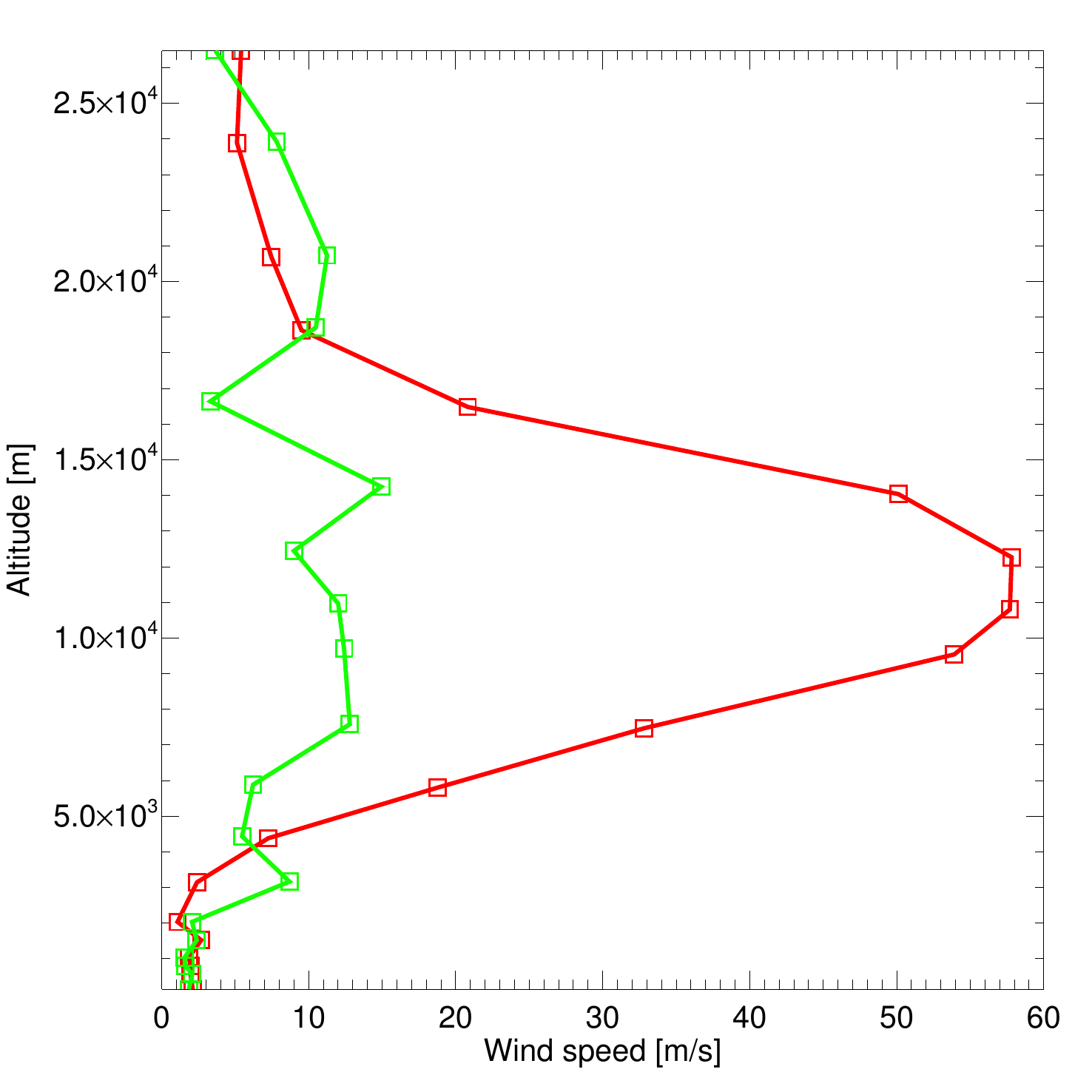}\includegraphics{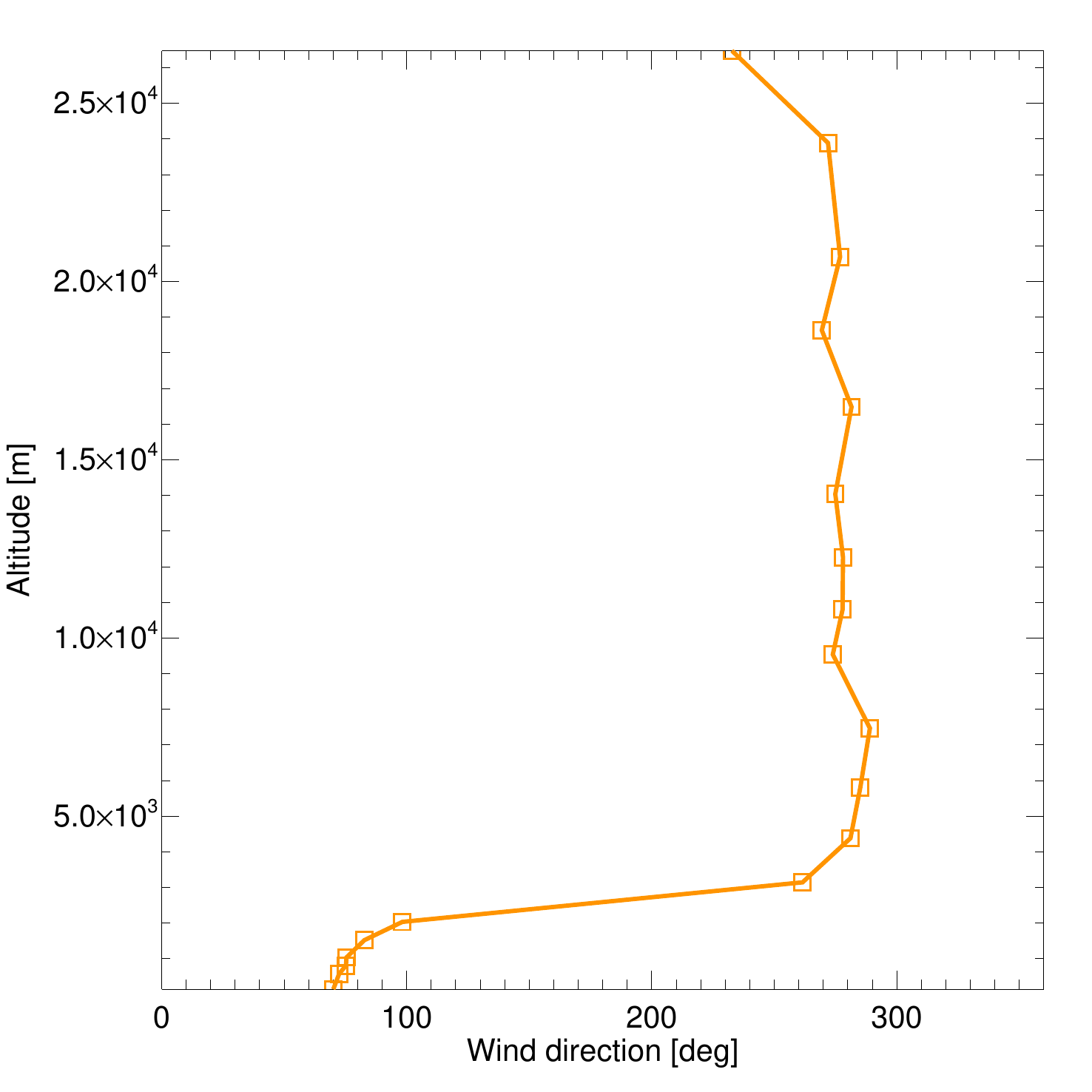}}
   \end{tabular}
   \end{center}
\caption{\label{fig:atmprof} Profiles of $C_n^2$ (left), wind speed (middle) and wind direction (right) used for the simulations. Two turbulence conditions are presented: with strong jet stream layers (red solid lines) and without jet stream layers (green solid lines). In both cases the same wind direction is used (orange solid line).}
\end{figure}

To produce the AO residual phase screens, we use two wind profiles showing a typical case with and without a strong jet stream. These wind profiles are real profiles measured at the Paranal observatory during operation on July 23$^{\mathrm{rd}}$ 2015, by the European Center for Medium-Range Weather Forecasts (ECMWF\footnote{For more information, visit: \url{https://www.ecmwf.int/}}). The wind profiles consist of $19$ layers distributed between $0.15~\mathrm{km}$ to $26.5~\mathrm{km}$ shown in Fig.~\ref{fig:atmprof}. Using the altitude measured by the ECMWF, the corresponding $Cn^2$ profiles have been interpolated from the 35-layered median $Cn^2$ profile for the Cerro Armazones site provided by ESO (ESO-258292). To simulate the AO residuals for both instruments, the same profiles are used, assuming a target star of magnitude $4.5$ in K-band and $4.7$ in I-band (similar to the star 51~Eri), located at a zenith angle of $30^{\mathrm{o}}$, and with a seeing of $0.65"$ (equivalent to $r_0(30^{\mathrm{o}})=84~\mathrm{cm}$ for METIS and $r_0(30^{\mathrm{o}})=21~\mathrm{cm}$ for SPHERE). From these profiles, the turbulence coherence times are therefore\cite{Sarazin2002} of about $11.4~\mathrm{ms}$ in the case of absent jet stream and about $2.1~\mathrm{ms}$ in the case of strong jet stream. %Note that the corresponding Greenwood frequency (the inverse of $\tau_0$) are of respectively $87~\mathrm{Hz}$ and $471~\mathrm{Hz}$. 

% and we chose a scalar gain of $0.45$ -> for METIS = 1.7ms et for SPHERE = 1.25ms
In the AO simulations, the frame delay is set to two so that the total delay of the AO loop is of $1.5~\mathrm{ms}$ for SPHERE and $2~\mathrm{ms}$ for METIS. We therefore expect to observe the WDH from a turbulence coherence time of about $2~\mathrm{ms}$ and below. In practice, at Paranal observatory, we clearly observe the WDH on SPHERE images as soon as the coherence time is below $2~\mathrm{ms}$. 

Table~\ref{tab:diff} gathers the fundamental differences between the VLT/SPHERE and ELT/METIS systems that have a significant impact on the AO-residuals. In the following we describe and simulate the effect of each of these differences on the wind driven halo. For both cases we considered a high-order DM with actuators distributed on a regular grid showing Gaussian influence functions. To simulate the high-contrast images from the AO residual phase screens, we considered the telescope pupil (including the central obstruction and spiders) and simulated images with an ideal coronagraph (without any undersized Lyot stop). In the ELT case, we ignored the segmentation of the primary mirror and considered a circular pupil in the inscribed circle of the actual ELT pupil. The simulated images are purely monochromatic at the observing wavelength. 
The resulting high-contrast images simulated for both systems under strong jet stream conditions are shown in Fig.~\ref{fig:intro}. These two images are obtained from a stack of $200$ instantaneous images generated from random AO residual phase screens. No photon noise is taken into account since the target star is bright enough and neither readout noise nor thermal background is taken into account.

\begin{table}[h]
\caption{Main differences between the VLT/SPHERE and ELT/METIS (as of the current design for the ELT and for the METIS SCAO) systems that are taken into account for the simulations to highlight the effect of each parameters to the wind driven halo.} \label{tab:diff}
\begin{center}       
\begin{tabular}{|l|l|l|}
\hline
\rule[-1ex]{0pt}{3.5ex} \textbf{Parameters} & \textbf{VLT/SPHERE} & \textbf{ELT/METIS}  \\
\hline
\rule[-1ex]{0pt}{3.5ex} Telescope diameter & $8~\mathrm{m}$  & $37~\mathrm{m}$ \\
\hline
\rule[-1ex]{0pt}{3.5ex} Number of DM actuators & $41 \times 41$ & $75 \times 75$  \\
\hline
\rule[-1ex]{0pt}{3.5ex} AO loop frequency & $1380~\mathrm{Hz}$ & $1000~\mathrm{Hz}$  \\
\hline
\rule[-1ex]{0pt}{3.5ex} Wavefront sensor type & Shack-Hartmann & Pyramid \\
\hline
\rule[-1ex]{0pt}{3.5ex} Wavefront sensing wavelength & I-band ($0.7~\mathrm{\mu m}$) & K-band ($2.20~\mathrm{\mu m}$) \\
\hline
\rule[-1ex]{0pt}{3.5ex} Imaging wavelength & H-band ($1.6~\mathrm{\mu m}$) & L-band ($3.7~\mathrm{\mu m}$)  \\
\hline 
\rule[-1ex]{0pt}{3.5ex} Pixel scale & $12.25~\mathrm{mas/px}$ & $5.47~\mathrm{mas/px}$  \\
\hline 
\end{tabular}
\end{center}
\end{table}

%------------------------------------------------------------
\subsection{Variation of the raw contrast in presence of WDH}
\begin{figure}[ht]
   \begin{center}
   \begin{tabular}{c} %% tabular useful for creating an array of images 
   \includegraphics[scale=0.5]{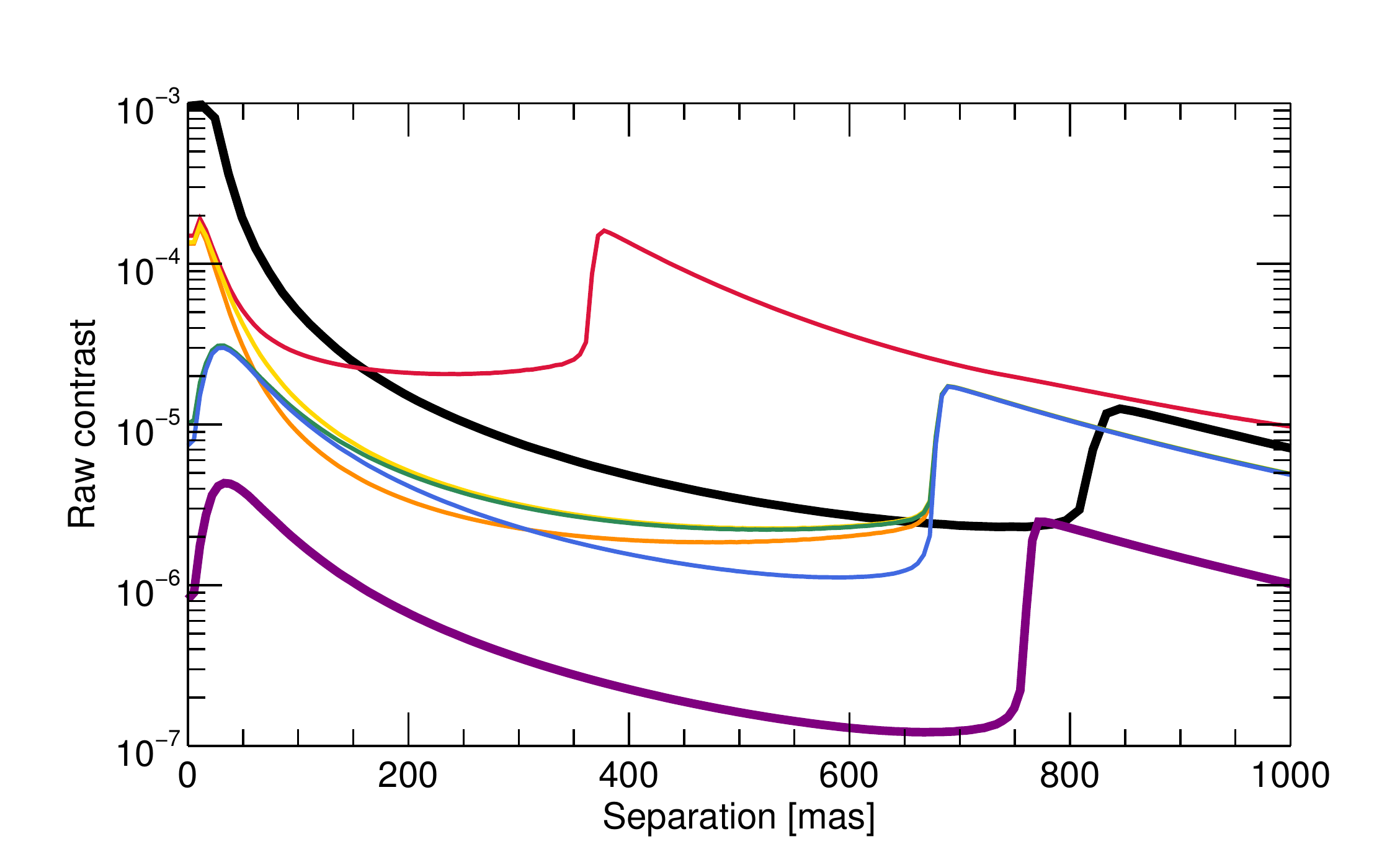}
    \begin{minipage}[c]{5cm}
    \centering
    \vspace{-6cm}
    \includegraphics[angle=-90,origin=c,scale=0.18]{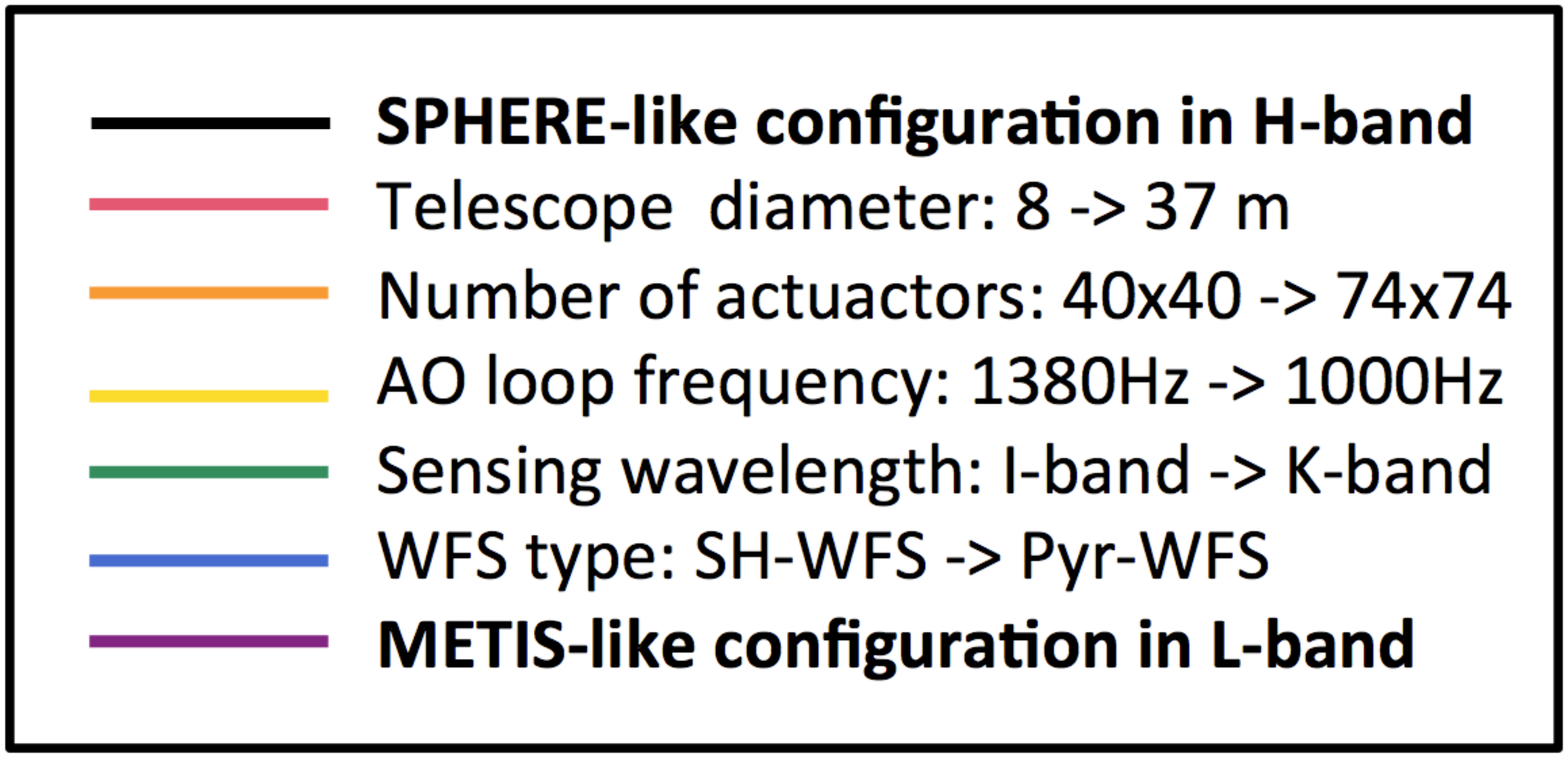}
   \end{minipage}
   \end{tabular}
   \end{center}
\caption{\label{fig:profilesall} Raw contrast obtained from simulations of infinite exposure using an ideal coronagraph when adding one-by-one the main differences between SPHERE (thick black line) and METIS (thick purple line). All simulations account for the same atmospheric turbulence profiles shown at Fig.~\ref{fig:atmprof} (case of strong jet stream, in red).}
\end{figure}

In this section we derive the raw contrast as a function of the separation to the star, obtained from ideal coronagraphic images of infinite exposure simulated by accounting only for the AO residuals under the conditions of strong jet stream. Starting from the SPHERE-like instrument working in the H-band, we changed the parameters one-by-one to see the effect of each key parameter, until we reach the METIS-like configuration. 
Figure.~\ref{fig:profilesall} shows the mean azimuthal profile for each case.
\begin{enumerate}
    \item When going from a telescope diameter of $8~\mathrm{m}$ to $37~\mathrm{m}$, as expected we reach a higher angular resolution, but the small number of actuators with respect to the size of the pupil create a dark hole at worse contrast (red line);
    \item When increasing the number of DM actuators (and sub-pupils of the WFS), we recover a better contrast at all separations and the cutoff frequency is almost located at the same physical separation (orange line);
    \item When decreasing the AO loop frequency, the contrast decreases as expected since the AO servolag error, creating the WDH, is higher (yellow line);
    \item When sensing at larger wavelength (from I to K-band), we suffer from less chromaticity hence the contrast increases at short separation (green line);
    \item When going from a Shack-Hartmann wavefront sensor (SH-WFS) to a Pyramid wavefront sensor (Pyr-WFS), we observe an increase of the contrast in the dark hole since the Pyr-WFS is more sensitive, shows smaller noise propagation and is less sensitive to aliasing (blue line);
    \item At last, when imaging at larger wavelength, from the H-band to the L-band, we observe a significant increase of the contrast in the dark hole since larger wavelengths are less affected by atmospheric turbulence (purple line).
\end{enumerate}

\begin{figure}[ht]
   \begin{center}
   \begin{tabular}{c} %% tabular useful for creating an array of images 
   \includegraphics[scale=0.65]{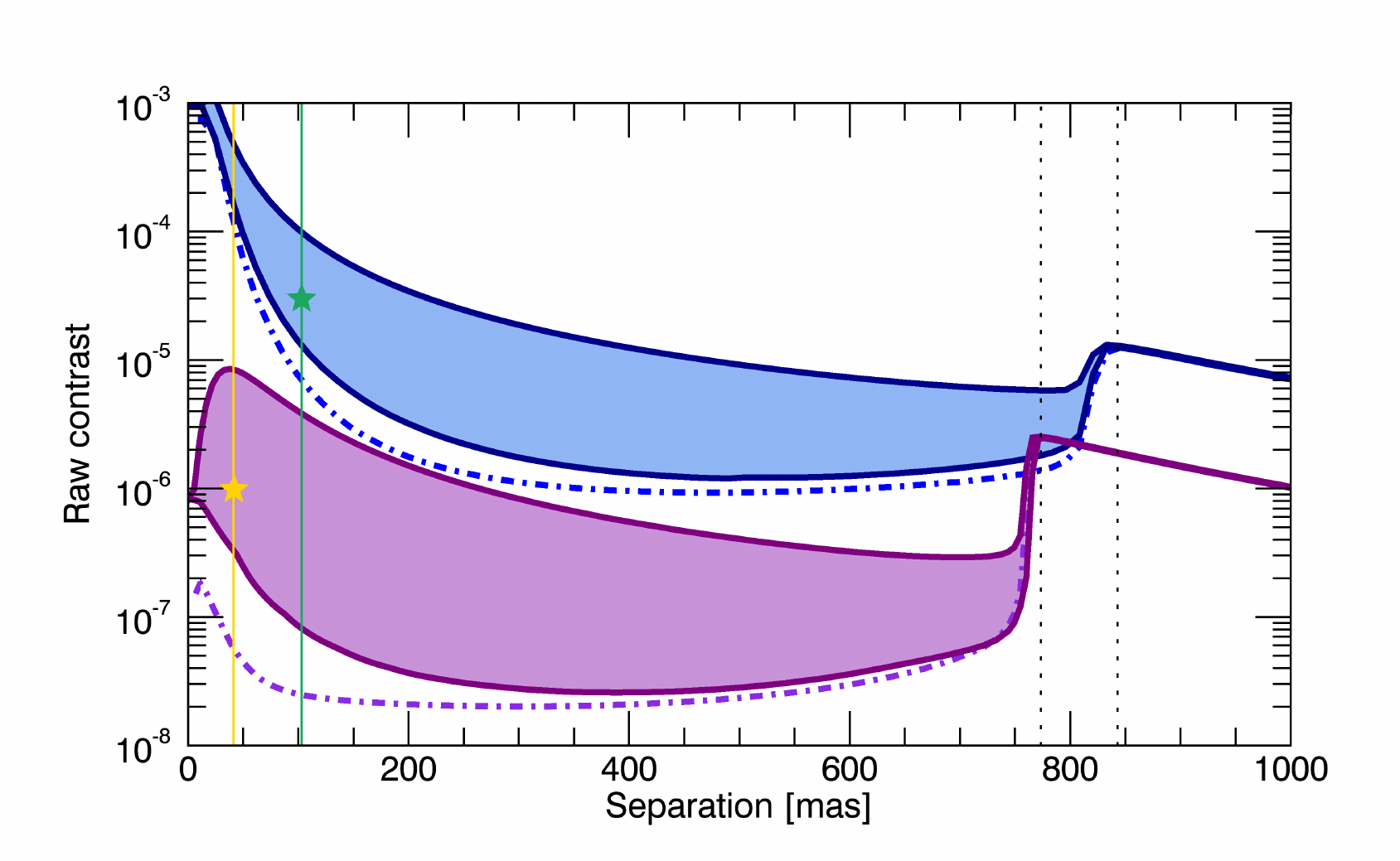}
   \begin{minipage}[c]{5cm}
    \centering
    \vspace{-6cm}
    \includegraphics[angle=-90,origin=c,scale=0.15]{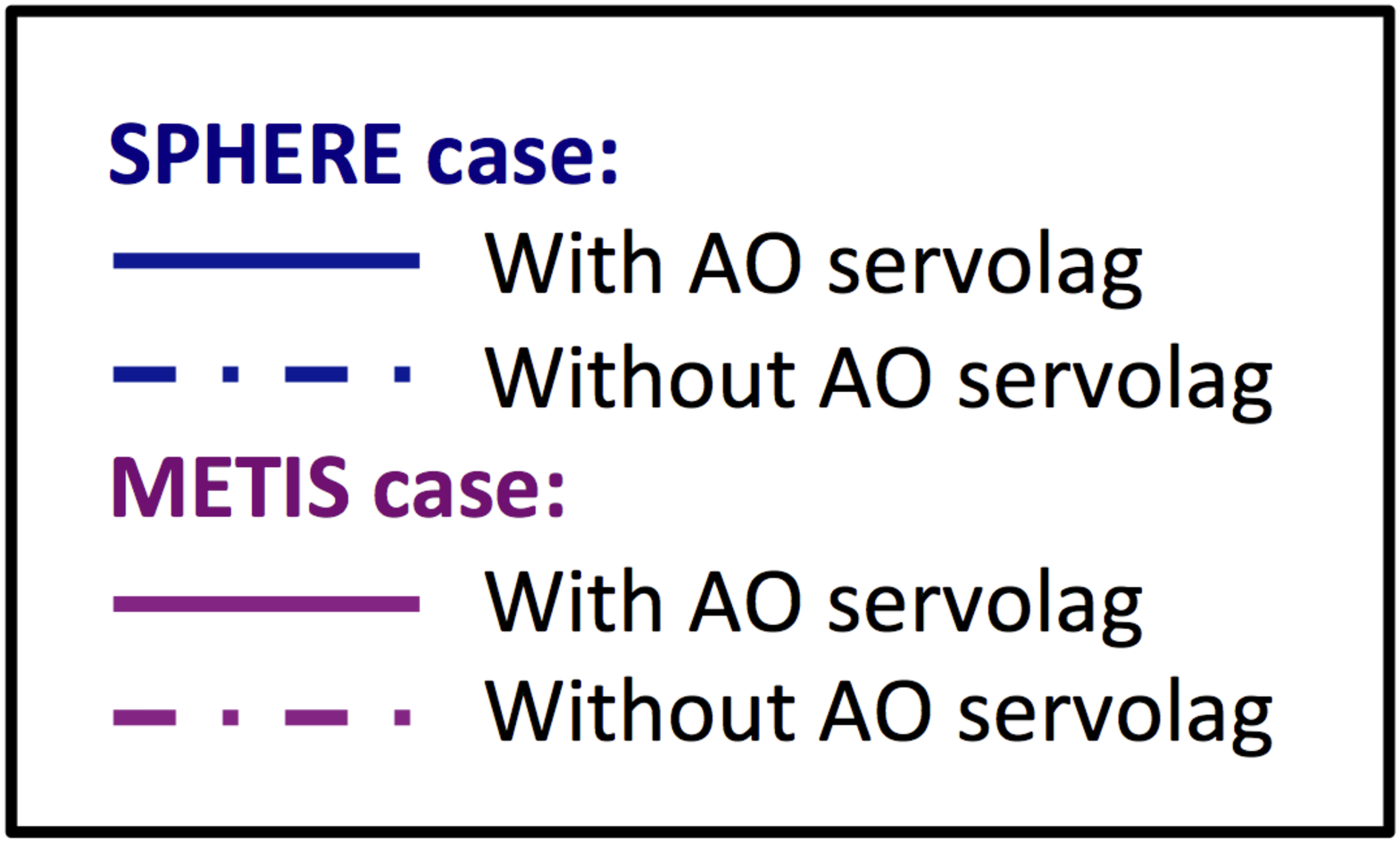}
   \end{minipage}
   \end{tabular}
   \end{center}
\caption{\label{fig:profilesccl} Comparison of the raw contrast obtained with an ideal coronagraph as a function of the separation for a SPHERE-like instrument (blue lines) and a METIS-like instrument (purple lines). In the ideal cases without AO servolag error (dash-dotted lines), the profiles are the azimuthal average of the image. In presence of AO servolag error, the shaded area shows the raw contrast from the azimuthal minimum (perpendicular to the WDH) to the azimuthal maximum (along the wind direction). The goal and requirement of METIS are highlighted (resp. green and yellow lines), as well as the cutoff frequencies (vertical dotted lines) for SPHERE ($843~\mathrm{mas}$ in H-band) and METIS ($773.5~\mathrm{mas}$ in L-band).}
\end{figure}

In that configuration, compared to SPHERE, for the same physical separation to the star, METIS gains slightly more than an order of magnitude in raw contrast. To highlight the effect of the AO servolag error in the raw contrast, Fig.~\ref{fig:profilesccl} shows the raw contrast profiles for SPHERE and METIS in the direction of the WDH and in the direction perpendicular to the WDH (solid lines), to be compared to the raw contrast reached without AO servolag error (dash-dotted lines). At small separations, the gain for SPHERE is not obvious since the AO of SPHERE is dominated by chromaticity and noise propagation. However for METIS, the gain is non-negligible when no AO servolag error is present (almost one order of magnitude in the case without WDH, to two order of magnitude with WDH). This further confirms that the AO servolag error is dominating the raw contrast budget at small angular separation for METIS. 

The requirement for METIS is to reach a post-processed contrast of $3.10^{-5}$ at $5\lambda/D$ ($103~\mathrm{mas}$) and the goal is to reach a post-process contrast of $1.10^{-6}$ at $2\lambda/D$ ($41~\mathrm{mas}$) in L-band. The AO servolag error is therefore not a limiting factor to reach the requirement, but it is a limiting factor for reaching the goals, unless we remove the WDH contribution, by either a better control scheme or by post-processing techniques. Overall, in terms of AO residuals, only the AO servolag error is an issue to reach the contrast goal for METIS, the other AO error terms are negligible to reach the goal.

%------------------------------------------------------------
\subsection{Amount of WDH in the focal plane images}
To account for the amount of WDH present in the images, we compare the power spectral densities (PSD) obtained with and without the AO servolag error, in the cases of METIS and SPHERE, for the two typical observing conditions with and without strong jet stream. Compared to the PSD, the WDH in the resulting focal plane image would be attenuated by the effect of the coronagraph (and is therefore coronagraph-dependent), but at separation larger than the coronagraph inner working angle, the PSD shows the absolute amount of light scattered into the WDH for a long exposure time. %In addition, other sources of error than AO residuals would scatter the starlight in the focal plane image (such as low wind effect, jitter or non-common path aberrations), at various spatial frequencies, that could attenuate further the WDH contribution. 

\begin{figure}[ht]
   \begin{center}
   \begin{tabular}{c} %% tabular useful for creating an array of images 
   \includegraphics[scale = 0.5]{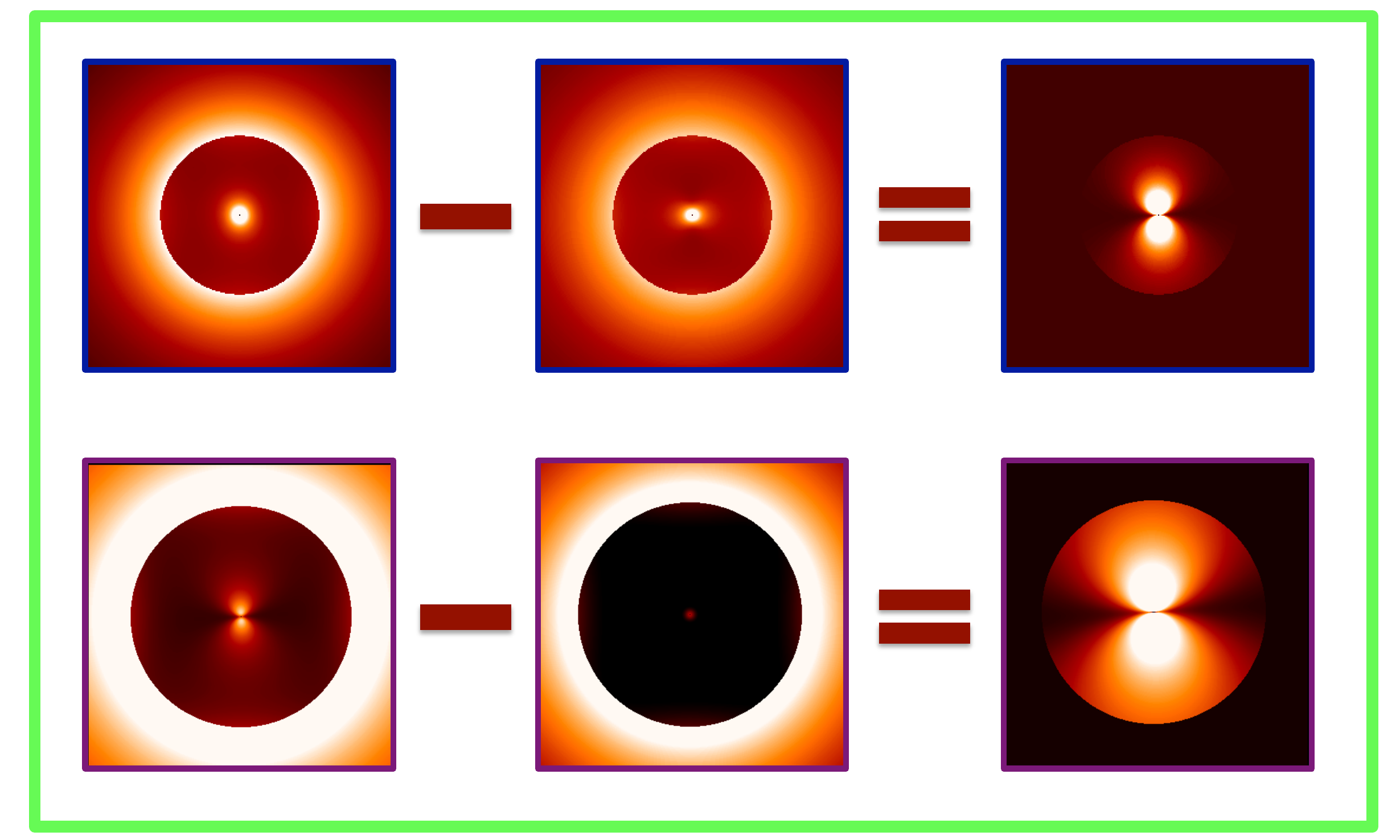}
   \\
   \includegraphics[scale = 0.5]{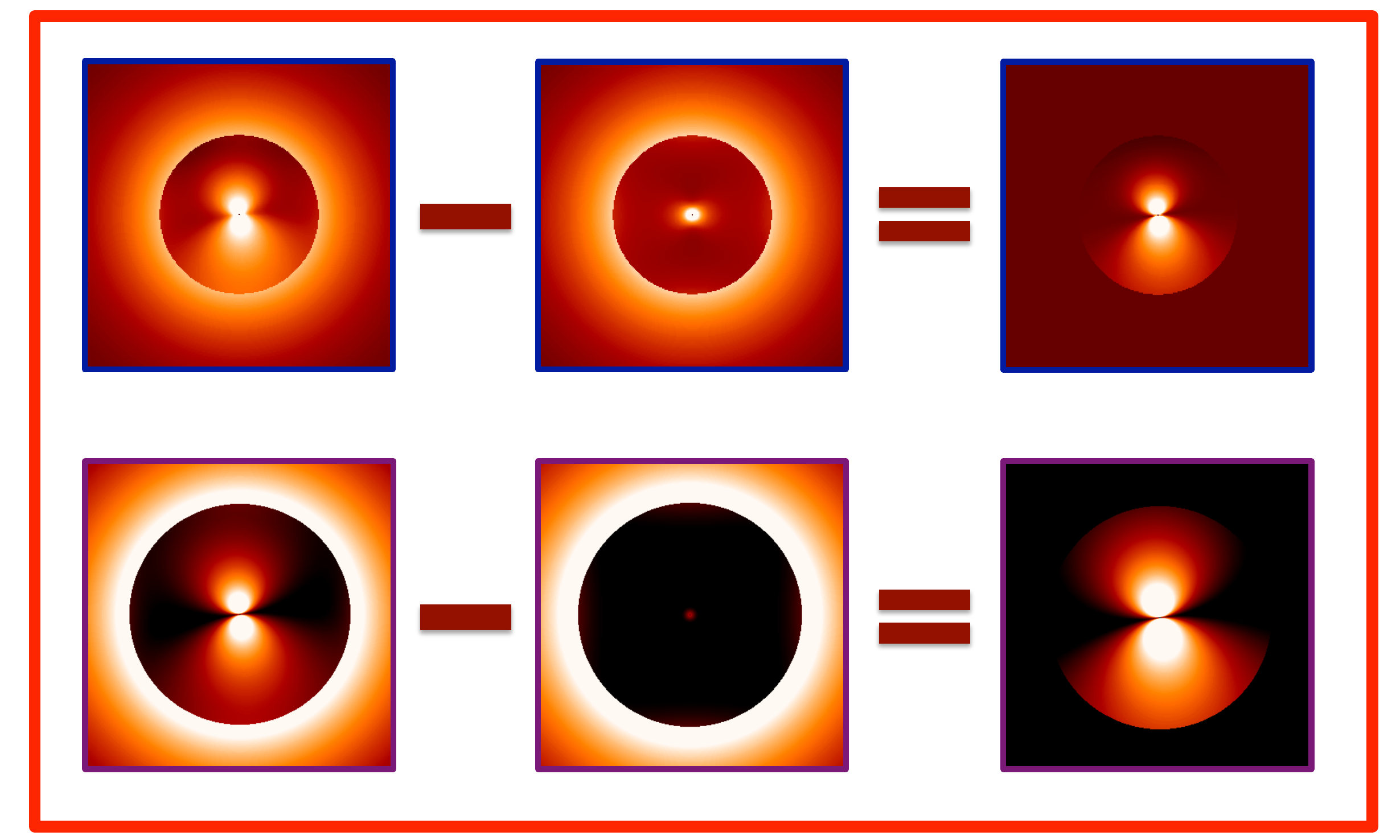}
   \end{tabular}
   \end{center}
\caption{\label{fig:psdjs}Power spectral densities after the AO correction for the cases without jet stream (green upper box) and with strong jet stream (red lower box). Left column shows the PSDs including the AO servolag error, the middle column shows the PSDs without the AO servolag error and the right column the difference of the two, that is to say the AO servolag error only. In each box, the upper row is the case for a SPHERE-like system (blue boxes) and the lower row is the case for a METIS-like system (purple boxes).} 
\end{figure}

Figure~\ref{fig:psdjs} shows the simulated PSD produced with weak (green upper box) and strong jet stream (red lower box) for both SPHERE-like (blue boxes) and METIS-like (purple boxes) systems. 
In the case of SPHERE, without the strong jet stream, $15.0\%$ of the starlight within the corrected area is produced by the AO servolag error (about $2.1\%$ of the starlight within the total field of view), and $69.7\%$ with the strong jet stream (about $21.5\%$ within the total field of view). 
In the case of METIS, without the strong jet stream, $51.3\%$ of the starlight within the corrected area is produced by the AO servolag error (about $0.6\%$ of the starlight within the total field of view), and $93.3\%$ with the strong jet stream (about $8.0\%$ of the starlight within the total field of view). 
The difference between the AO servolag contribution with respect to the total light and its contribution in the corrected area only is due to the fact that for METIS, the sub-pupils are $50~\mathrm{cm}$ wide whereas they are $20~\mathrm{cm}$ wide for SPHERE, making the DM fitting error a much higher contribution in the METIS case.

In the specific case without strong jet stream, the AO servolag signature is much more visible in the METIS case and is hidden by other dominant terms in the SPHERE case, as visible in Fig.~\ref{fig:psdjs}. 

In the case with strong jetstream, the AO servolag signature is indeed along the direction of the wind at the jet stream layer shown in Fig.~\ref{fig:atmprof}. 
As expected, over the total amount of light, the WDH is less prominent in the METIS case compared to the SPHERE case, when taking into account all of the differences between both systems. The larger diameter of the telescope, the larger sensing wavelength and the use of a pyramid WFS do decrease the AO servolag error (and therefore the WDH), even though the AO loop speed is slightly slower for METIS than for SPHERE. In the corrected area however, the amount of WDH is higher for METIS than for SPHERE because the AO servolag error is dominant with respect to other AO errors in the case of METIS, which suffers less from chromaticity, anisoplanetism, noise propagation and aliasing. Note that these simulations do not take into account exogenous errors such as pupil fragmentation\cite{Bonnet2018SPIE}, vibrations, calibration errors, non-common path aberrations or low wind effect, and are purely monochromatic. 

%------------------------------------------------------------
\subsection{Asymmetry of the WDH in the focal plane images}
As previously, we computed the absolute amount of asymmetry directly using the simulated PSD that are shown in Fig.~\ref{fig:psdjs}. We define the asymmetry factor as the difference between the light contained in the bright wing and the light contained in the faint wing, divided by the total amount of light in the two wings\cite{Cantalloube2018}. An asymmetry factor of $100\%$ is therefore a purely asymmetric WDH, showing only one wing, whereas an asymmetry factor of $0\%$ is a purely symmetric WDH. 

In the case of SPHERE, without the strong jet stream, other errors (such as anisoplanetism and chromaticity) are dominating over the servolag and therefore no asymmetry can be observed. With strong jet stream, the asymmetry is of $21.1\%$. 
In the case of METIS, as shown previously, even without strong jet stream, the AO servolag error dominates over other sources of error but the asymmetry is only of $1.4\%$. With strong jet stream, the asymmetry is of $21.3\%$, similar to that of SPHERE.

As explained in Sect.~\ref{sec:wdh}, from SPHERE to METIS, we expect the asymmetry to decrease due to the larger telescope diameter but to increase due to the slower AO loop and larger imaging wavelength. All in all, the fraction of asymmetry is similar for a SPHERE-like system in H-band and a METIS-like system in L-band when enough WDH is visible in the images. However the Cerro Armazones that will host the ELT is higher than Cerro Paranal hosting the VLT. We therefore expect the ELT to undergo less scintillation and therefore show less asymmetry.

%%%%%%%%%%%%%%%%%%%%%%%%%%%%%%%%%%%%%%%%%%%%%%%%%%%%%%%%
\section{Conclusions}
The wind driven halo (WDH) is a butterfly-shaped structure observed in high-contrast images provided by the latest generation of instruments dedicated to the detection and characterisation of exoplanets and circumstellar disks. This structure shows up at a contrast of about $10^{-4}$ and is limiting the contrast performance of the instrument, especially for imaging extended sources, for which it is not possible to apply a spatial high-pass filter. The WDH is due to the limited temporal bandwidth of the AO correction and appears when the turbulence coherence time is lower than the absolute AO loop delay. 
From simulations of power spectral densities of AO residuals, using an analytical AO simulator and realistic turbulence conditions with strong jet stream, we computed the amount of starlight within the corrected area that belongs to the WDH: about $70\%$ for the VLT/SPHERE instrument, consistent to what is observed on real on-sky data, and about $93\%$ for the ELT/METIS instrument, since the AO servolag is the dominant AO error term. %with respect to other AO error terms. 

The asymmetry of the WDH is due to the interference between the delayed phase error (introduced by the AO servolag error) and amplitude errors (introduced by scintillation). From simulations of the power spectral densities of AO residuals, we computed the amount of asymmetry of the WDH: about $20\%$ for the VLT/SPHERE instrument, consistent to what is observed on real on-sky data, and also about $20\%$ for the ELT/METIS instrument, since the parameters at play are balancing at the end.

Currently, the post-processed contrast performance of SPHERE is limited by the WDH that has an occurrence rate ranging from $30$ to $40\%$ of the observation time, according to the seasonal occurrence of strong jet stream above Paranal observatory. 
For ELT/METIS, the AO error budget is dominated by the AO servolag error and, under short turbulence coherence time, this will significantly affect the contrast performance in a preferential direction, even after applying post-processing techniques. This effect can be alleviated by specific AO control scheme, such as predictive control, or by post-processing techniques adapted to deal with the WDH. Furthermore, when assessing the performance of the instrument in terms of contrast curves, as it is usually done in the high-contrast community, one should provide not only the azimuthal average of the residuals, but also the minimum and maximum contrast reached as a function of the separation to the star to account for the WDH.

%%%%%%%%%%%%%%%%%%%%%%%%%%%%%%%%%%%%%%%%%%%%%%%%%%%%%%%%
\section{Perspectives}
For high-contrast instruments currently in operation, there are various ways to upgrade the instrument in order to reduce the WDH in the final images. One way is to significantly decrease the AO-loop delay, which requires a faster wavefront sensor detector, hence a more sensitive wavefront sensor, along with a faster real-time computer and faster deformable mirror response. In that framework, SPHERE+, an upgrade of SPHERE, has been recently defined and proposed at the VLT-2030 conference (June 2019, Garching, Germany) and GPI~2.0\cite{Chilcote2018gpi2} has been proposed and will be commissioned at the Gemini-North telescope in 2022. Another way is to make use of predictive control algorithms, based for instance on machine learning techniques\cite{males2018ground,jensen2019demonstrating}. This solutions is already routinely used at the Subaru/SCeXAO instrument\cite{sahoo2018subaru}. At last, dedicated post-processing techniques could be developed to tackle this specific signature.

As for the ELT/METIS instrument, the present study can be refined using more atmospheric profiles and more realistic AO simulations, in order to assess how much the WDH affects the contrast under various observing conditions. This study will have a direct impact on how to optimize the observing schedule to not be limited by this effect. %, (Working in the thermal infrared, METIS will be used when observing conditions show a low amount of water vapor).  
On the SCAO design side, control algorithms to optimize the AO loop delay are currently being defined and predictive control solutions are being explored\cite{van2017predcontelt}.

A similar study can be made for any other instrument foreseen to be installed on giant segmented mirror telescopes (GSMT), using the turbulence parameters of the chosen observing site. 

A last aspect worth mentioning, is the effect of the WDH in PSF reconstruction outside of the HCI field. In the context of ELT instruments, we are currently on the way to reach such a precision in PSF reconstruction that the elongation provoked by the WDH, even without coronagraph device, might be present in the image and slightly bias high precision measurements such as astrometric measurements\cite{Beltramo2019prime}.

%%%%%%%%%%%%%%%%%%%%%%%%%%%%%%%%%%%%%%%%%%%%%%%%%%%%%%%%
%\acknowledgments % equivalent to \section*{ACKNOWLEDGMENTS}       

%%%%%%%%%%%%%%%%%%%%%%%%%%%%%%%%%%%%%%%%%%%%%%%%%%%%%%%%
\bibliography{main} % bibliography data in report.bib
\bibliographystyle{spiebib3} % makes bibtex use spiebib.bst

\end{document}